\documentclass[11pt,a4paper]{article}

\usepackage{amsmath}
\usepackage{amsthm,amssymb}

\usepackage{graphicx} 

\newcommand{\beq}{\begin{equation}}
\newcommand{\eeq}{\end{equation}}

\numberwithin{equation}{section}

\begin{document}
\title{Exact Solution of the Discrete (1+1)--dimensional RSOS Model in
  a Slit with Field and Wall Interactions}
\author{A L Owczarek$^1$ and T Prellberg$^2$\\
  \footnotesize
  \begin{minipage}{13cm}
    $^1$ Department of Mathematics and Statistics,\\
    The University of Melbourne, Parkville, Vic 3010, Australia.\\
    \texttt{owczarek@unimelb.edu.au}\\[1ex] 
$^2$ School of Mathematical Sciences\\
Queen Mary University of London\\
Mile End Road, London E1 4NS, UK\\
\texttt{t.prellberg@qmul.ac.uk}
\end{minipage}
}

\maketitle  

\begin{abstract}

  We present the solution of a linear Restricted Solid--on--Solid
  (RSOS) model confined to a slit. We include a field-like energy,
  which equivalently weights the area under the interface, and also
  include independent interaction terms with both walls. This model
  can also be mapped to a lattice polymer model of Motzkin paths in
  a slit interacting with both walls and including an osmotic
  pressure.

  This work generalises previous work on the RSOS model in the
  half-plane which has a solution that was shown recently to exhibit a
  novel mathematical structure involving basic hypergeometric
  functions ${}_3\phi_2$. Because of the mathematical relationship
  between half-plane and slit this work hence effectively explores the
  underlying $q$-orthogonal polynomial structure to that solution.  It
  also generalises two other recent works: one on Dyck paths weighted
  with an osmotic pressure in a slit and another concerning Motzkin
  paths without an osmotic pressure term in a slit.

\end{abstract}

\section{Introduction}

 Solid--on--Solid (SOS) models describe the interface between
 low-temperature phases, originally in magnetic systems such as
 Ising-like models \cite{te52,di88}, though now more generally. They are effectively
   directed models in $d + 1$ dimensions.  The configurations involved 
in the linear (1+1)--dimensional case, modelling the interface in a two-dimensional 
system, have also been used to model the backbone of a polymer in
solution \cite{prsv89}. 
The critical phenomena associated with this model describe wetting transitions of
the interface with a wall \cite{di88}.
For the SOS model the phase diagram contains a wetting transition at
finite temperature $T_w$ for zero field and complete wetting occurs taking
the limit $H\to0$ for $T\geq T_w$ \cite{owczarek93a}.  The basic model
is naturally described in the half-plane but it is also natural to
describe it in the confined geometry of the slit.

The linear SOS model with magnetic field and wall interaction was
solved in \cite{owczarek93a}.  Recently the restricted SOS (RSOS),
where the interface takes on a restricted subset of configurations,
was solved with the same interactions of field and single wall
interaction in the half-plane \cite{op09}. This has proven to be
mathematically quite interesting as both the method of solution and
the functions involved were novel. It was found that the solution could
be expressed as ratios of linear combinations of terms involving the 
basic hypergeometric function ${}_3\phi_2$. Also
recently the polymer models of Dyck paths \cite{borw05} and Motzkin
paths \cite{birw07} in a slit with separate interactions with both
surfaces have been considered, without field-like terms. Here the
solutions in the slit prove interesting both mathematically and
physically. They are of interest physically because the infinite slit
limit was shown to be subtly different to the half plane, realising a
separate phase diagram \cite{borw05}. Mathematically the slit exposes
the orthogonal polynomial structure of the problem and uncovers hidden
combinatorial relationships \cite{birw07}. Finally, Dyck paths in a
slit with wall interactions and weighted by the area under the path,
equivalent to a field term in the SOS models, have only recently been
analysed \cite{bop09}, and show a rich $q$-orthogonal polynomial
structure.  To explore further this area of research here we consider
the RSOS model in a slit geometry with both separate wall interactions and a
field/osmotic pressure term in the energy. We derive the novel
$q$-orthogonal polynomials for this problem which give us the exact
solution of the generating function.

\section{The RSOS model in a slit}

The RSOS model we analyse can be described as follows. Consider a two-dimensional square lattice 
in a slit of width (or thickness) $w\geq 1$. For each column $i$ of the
surface a segment of the interface is placed on the horizontal link at
height $0 \leq r_i\leq w $, and
successive segments are joined by vertical segments to form a partially directed interface
with no overhangs. The configurations are given the energy
\beq
-\beta E=-K\sum_{i=1}|r_i-r_{i-1}|-H\sum_{i=1} r_i+B_0 \sum_{i=1}\delta_{r_i,0} + B_w\sum_{i=1}\delta_{r_i,w}\; .
\eeq

\begin{figure}[ht]
\begin{center}
\includegraphics[width=0.9\textwidth]{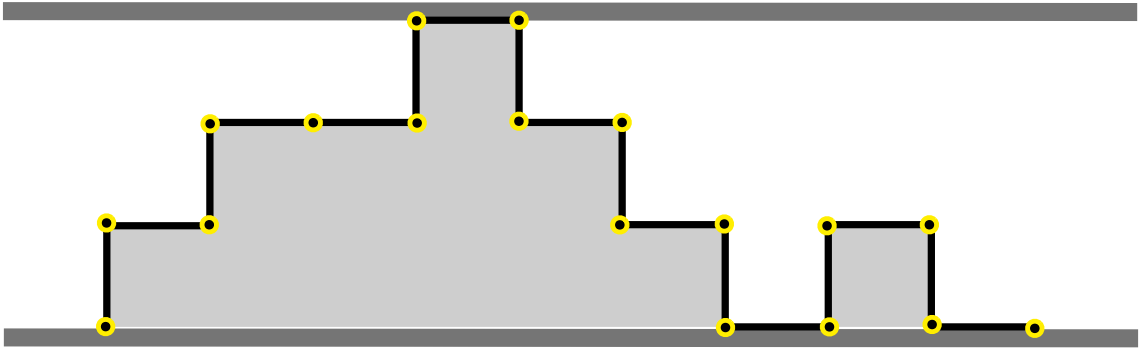}
\caption{A typical RSOS configuration beginning on the surface and finishing on the surface with
a horizontal step: each horizontal step is assigned a weight $x$, each vertical step a weight $y$,
each unit of area a weight $q$, each step that touches the bottom
surface an additional weight $a$ while each step that touches the top
surface an additional weight $b$.
The length of the configuration shown here is $N=9$ in a  slit of
width $w=3$, and the heights are $r_0=0$, $r_1=1$, $r_2=r_3=2$, $r_4=3$, 
$r_5=2$, $r_6=1$, $r_7=0$, $r_8=1$, and $r_9=0$. The weight of this
configuration equals $x^9y^8q^{12}a^2 b$.}
\label{figure1}
\end{center}
\end{figure}

As in \cite{op09}, we discuss the RSOS model in terms of lattice paths.
An RSOS path is a partially directed self-avoiding path with no steps into
the negative $x$-direction and no successive vertical steps. To be precise,
an RSOS path of length $N$ with heights $r_0$ to $r_N$ has horizontal steps at heights
$r_1,\ldots,r_N$, and vertical steps between heights $r_{i-1}$ and $r_i$ for $i=1,\ldots,N$,
but no horizontal step associated with $r_0$. This means that an RSOS path starts at height $r_0$ 
with either a horizontal step (if $r_1=r_0$) or vertical step (if $r_1\neq r_0$), but must 
end at height $r_N$ with a horizontal step. Figure \ref{figure1} shows an example.

The partition function for the RSOS paths of length $N$ in a slit of
width $w\geq 1$ with ends fixed at heights
$r_0\geq0$ and $r_N\geq0$, respectively, is given by
\beq
Z^w_1(r_0;r_1)=\begin{cases}\exp(-\beta E(r_0;r_1))\;,&|r_0-r_1|\leq 1\\ 0\;, &|r_0-r_1|>1\;,\end{cases}
\eeq
and
\beq
Z^w_N(r_0;r_N)=\sum_{\substack{0\leq r_1,\ldots,r_{N-1}\leq w\\|r_i-r_{i-1}|\leq1}}
\exp(-\beta E(r_0;r_1,\ldots,r_N))\;,
\qquad N=2,3,\ldots \; ,
\eeq
where 
\beq
-\beta E(r_0;r_1,\ldots,r_N)=
-K\sum_{i=1}^N|r_i-r_{i-1}|-H\sum_{i=1}^Nr_i+B_0\sum_{i=1}^N\delta_{r_i,0}
+B_w\sum_{i=1}^N\delta_{r_i,w}
\; .
\eeq
Here, we shall consider paths with both ends attached to the surface, i.e. we
shall focus on the partition function 
\beq
Z^w_N=Z^w_N(0;0)\;.
\eeq
We define
\beq
y=\exp(-K)\;,\quad q=\exp(-H)\;,\quad
a=\exp(B_0)\;,\quad
\mbox{and}\quad
  b=\exp(B_w)\;,
\eeq
so $y$ is a temperature--like, $q$ a magnetic field--like and $a$ and $b$ are
binding energy--like variables,
and write
\beq
Z^w_N=Z^w_N(y,q,a,b)\;.
\eeq
The (reduced) free energy is then 
\beq
\kappa(w;y,q,a,b)=\lim_{N\to\infty}\frac1N\log Z^w_N(y,q,a,b)\;.
\eeq
Define the generalised (grand canonical) partition function, or
simply generating function, as
\beq
\label{partition_function}
G_w(x,y,q,a,b)=1+\sum_{N=1}^\infty x^N Z^w_N(y,q,a,b)\;.
\eeq
Thus, the radius of convergence $x_c(w;y,q,a,b)$ of $G^w(x,y,q,a,b)$ with 
respect to the series expansion in $x$ can be identified as
$\exp(-\kappa(w;y,q,a,b))$, hence
\beq
 \kappa(w;y,q,a,b) = - \log x_c(w;y,q,a,b)\;.
\eeq

It is convenient to consider $G$ as a combinatorial generating function for 
RSOS paths, where $x$, $y$, $q$, $a$ and $b$ are counting variables for 
appropriate properties of those paths. Interpreted in such a way, $x$ and $y$ 
are the weights of horizontal and vertical steps, respectively, $q$ 
is the weight for each unit of area enclosed by the RSOS path and the $x$-axis, 
$a$ is  an additional weight for each step that touches the bottom
surface while $b$ is  an additional weight for each step that touches the top
surface.
For example, the weight of the configuration in Figure \ref{figure1}
is $x^9y^8q^{12}a^2 b$.

If we send $w\rightarrow\infty$ then we recover the generating
function of the half-plane
\beq
G^{hp}(x,y,q,a) = 1+\sum_{N=1}^\infty x^N Z^{hp}_N(y,q,a)
\eeq
where $Z^{hp}_N(y,q,a) = Z^w_N(y,q,a,b)$ for $w>N$, noting that the
paths can have no more vertical steps than horizontal steps in an RSOS
path. 

We find easily the first few terms of $G^{hp}$ as a series expansion in $x$,
\beq
G^{hp}(x,y,q,a)=1+ a x+(a^2+ a y^2q)x^2+\ldots\;,
\eeq
where the constant term corresponds to a zero-step path starting and 
ending at height zero with weight one.

Note that the radius of convergence of the half-plane generating
function $x^{hp}_c(y,q,a,b)$ is not a priori the limit of the slit
$x_c(w;y,q,a,b)$ as was demonstrated in \cite{borw05} for the
corresponding Dyck path problem with $q=1$.

\section{Exact solution for the generating function}

\begin{figure}[ht]
\begin{center}
\includegraphics[width=\textwidth]{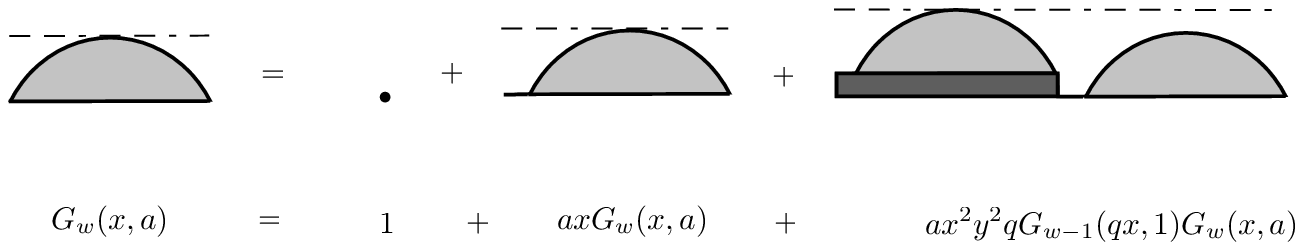}
\caption{The diagrammatic form of the functional equations for RSOS paths, indicating the
combinatorial decomposition of RSOS paths.}
\label{figure2}
\end{center}
\end{figure}

The key to the solution is a combinatorial decomposition of RSOS paths which leads to a
functional equation for the generating function $G_w$. This decomposition is done with
respect to the left-most horizontal step touching the surface at height zero, and is
shown diagrammatically in Figure \ref{figure2}.  

We distinguish three cases:
\begin{itemize}
\item[(a)] The RSOS path has zero length, and there is no horizontal step at height zero. The
contribution to the generating function is $1$.
\item[(b)] The RSOS path starts with a horizontal step, which therefore is at height zero. The rest of
this path is again a RSOS path. The contribution to the generating function is 
$\kappa xG_w(x,y,q,\kappa)$.
\item[(c)] The RSOS path starts with a vertical step. Then there will be a left-most horizontal step 
at height zero, and removing this step cuts the path into two pieces. The left path starts with a vertical 
and horizontal step, followed by an RSOS path starting and ending at height one and not touching the
surface, followed by a vertical step to height zero. This left path is
effectively in a slit of width $w-1$. The right path is again a RSOS
path in the slit of width $w$. The
contribution to the generating function is $yxqG_{w-1}(qx,y,q,1)y\kappa xG_w(x,y,q,\kappa)$.
\end{itemize}
Put together, this decomposition leads to a functional-recurrence equation for the generating function 
\beq
\label{functeq}
G_w(x,y,q,a,b)=1+ a xG_w(x,y,q,a,b)+ a qx^2 y^2
G_{w-1}(qx,y,q,1,b)G_w(x,y,q,a,b)\;,
\eeq
with ``initial condition''
\beq
G_0(x,y,q,a,b) = \frac{b}{1-abx}\;.
\eeq
If we send $w\rightarrow\infty$ then we recover the functional equation of
the half-plane
\beq
\label{hpfuncteq}
G^{hp}(x,y,q, a)=1+a x G^{hp}(x,y,q, a)+a qx^2y^2G^{hp}(qx,y,q,1)G^{hp}(x,y,q,a)\;.
\eeq
whose solution is given in \cite{op09}.

We can rewrite (\ref{functeq}) as
\beq
\label{recursion}
G_w(x,y,q,a,b)=\frac{1}{1- a x - a qx^2 y^2 
G_{w-1}(qx,y,q,1,b)}\;,
\eeq
For general values of $q$, a formal iteration of (\ref{recursion}) leads to
a continued fraction expansion
\beq
\label{confrac}
G_w(x,y,q,a,b)=
\cfrac{1}{1-a x-
\cfrac{a qx^2y^2}{1-qx-
\cfrac{q^3x^2y^2}{1-q^2x-
\cfrac{q^5x^2y^2}{1-q^3x-
\cfrac{q^7x^2y^2}{
 \genfrac{}{}{0pt}{}{}{\ddots 
            \genfrac{}{}{0pt}{}{}{
              \genfrac{}{}{0pt}{}{}{ 
              \genfrac{}{}{0pt}{}{}{ 
                \cfrac{}{1 - q^{w-1} x -
                  \cfrac{b q^{2w-1} x^2 y^2}{1-bq^{w} x }}}}}}}}}}}\;. 
\eeq

However, there is a non-trivial method to solve the functional
equation for $G$ as a ratio of $q$-orthogonal polynomials. 
It is clear that the generating function can be written as a
rational function
\begin{equation}
\label{Lrational1}
G_w(x,y,q,a,b)= \frac{P_w(x,y,q,a,b)}{Q_w(x,y,q,a,b)}
\end{equation}
though it does not simply follow to write expressions for these.
It does however follow from the theories of continued fractions and orthogonal
polynomials (see pages 256--257 of Andrews, Askey and Roy \cite{andrews1999a-a})
that both the numerator $P_w$ and denominator $Q_w$ of the generating
function satisfy recursions
\begin{equation}
\label{Precurrence}
P_w(x,y,q,a,b) = \begin{cases} b\;, & w=0\\
                              1- b qx\;, & w=1\\
                             (1-b q^w x)P_{w-1}(x,y,q,a,1) - b q^{2w-1}x^2 y^2
                             P_{w-2}(x,y,q,a,1)\; & w\geq 2
\end{cases}
\end{equation}
and 
\begin{equation}
\label{Qrecurrence}
Q_w(x,y,q,a,b) = \begin{cases} 1 - ab x\;, & w=0\\
                             1 - b q x - ax(1-bqx(1-y^2))\;, & w=1\\
                             -b q^w x Q_{w-1}(x,y,q,a,1) - b q^{2w-1}x^2 y^2
                             Q_{w-2}(x,y,q,a,1)\; & w\geq 2\;.
\end{cases}
\end{equation}
So it is the functions $Q_w(x,y,q,a,b)$ that are the $q$-orthogonal
polynomials referred to in the Introduction.

One can immediately note that
\begin{equation}
\label{den-num-relationship}
P_w(x,y,q,a,b) = Q_w(x,y,q,0,b)
\end{equation}
so that 
\begin{equation}
\label{Grational2}
G_w(x,y,q,a,b)= \frac{Q_w(x,y,q,0,b)}{Q_w(x,y,q,a,b)}\;.
\end{equation}

We now form the width generating function for the denominator as
\begin{equation}
\label{width-gen-fn}
W(t,x,y,q,a,b) = \sum_{w=0}^{\infty} Q_w(x,y,q,a,b) t^w
\end{equation}
and find a functional equation for $W(t)$ from the recurrence
(\ref{Qrecurrence}) as
\begin{align}
\label{funct-eqn-W}
W(t,x,y,q,a,b) = 1 &- a b x - a b q x^2 y^2 t 
+ t W(t,x,y,q,a,1)\\ 
&- b q x t W(qt,x,y,q,a,1) - b q^3 x^2 y^2 t^2 W(q^2 t,x,y,q,a,1) \nonumber
\end{align}
Unlike in the case of Dyck paths in a slit \cite{bop09}, one cannot
solve for $W(t)\equiv W(t,x,y,q,a,1)$ by direct iteration, as the functional equation
involves $W(t)$, $W(qt)$, and $W(q^2t)$.
So let us return to $Q_w(x,y,q,a,b)$ and consider the case $b=1$, as this is
all that is required to find the full solution. Let 
\beq
\label{defnR}
R_w \equiv Q_w(x,y,q,a,1)\;.
\eeq
From the recurrence (\ref{Qrecurrence}) we have for $w\geq 2$ that
\begin{equation}
\label{Rrecurrence}
R_w =   (1- q^w x) R_{w-1}-  q^{2w-1}x^2 y^2 R_{w-2}
\end{equation}
with initial conditions
\begin{align}
\label{Rinitial}
R_0 =&   \; 1 - a x                    \;,  \\
R_1 = &  \; 1 - q x - ax(1-bqx(1-y^2))\;. \nonumber 
\end{align}

Attempting to mimic aspects of the half-plane solution \cite{op09}, we
define $S_w$ via
\beq
\label{Ransatz}
R_w= (-1)^w q^{w(w+1)/2} x^w S_w\;.
\eeq
This gives us the recurrence
\begin{equation}
\label{Srecurrence}
q^wx(S_w  - S_{w-1} +   y^2 S _{w-2}) + S_{w-1}=0
\end{equation}
with initial conditions
\begin{align}
\label{Sinitial}
S_0 =& \;  1 - a x                    \;, \\
-q x S_1 =& \;  1 - q x - ax(1-bqx(1-y^2))\;. \nonumber 
\end{align}
Continuing with inspiration from the half-plane solution \cite{op09},
we try the Ansatz
\beq
\label{Sansatz}
S_w=\mu^w \sum_{n=0}^\infty c_n q^{-nw}\;.
\eeq
For $n=0$ we have
\beq
\label{cinit}
x(\mu^2-\mu+y^2) c_0 =0
\eeq
and
for $n>0$ we have
\beq
c_n = - \frac{\mu q^n c_{n-1}}{(y^2 q^{2n}- \mu q^n + \mu^2)qx}
\eeq
This implies that $\mu$ satisfies
\beq
\label{mu-eqn}
\mu^2-\mu+y^2 =0
\eeq
for our Ansatz to work.
We now parametrise $y$ via
\beq
\label{lambda}
y^2 = \lambda(1-\lambda)
\eeq
which changes (\ref{cinit}) into
 \beq
\label{cinit2}
x (\mu-\lambda)(\mu-1 +\lambda) c_0 =0
\eeq
and the recurrence into
\beq
c_n =  -\frac{\mu q^n c_{n-1}}{(\mu-\lambda q^n)(\mu-(1-\lambda)q^n)qx}
\eeq
We see immediately that either $\mu=\lambda$ or $\mu=1-\lambda$ and
that we  have two solutions from our Ansatz. This leads to the general
solution for $S_w$ as
\begin{align}
\label{Ssoln}
S_w = A \lambda^w &\sum_{n=0}^\infty 
\left(-\frac{1}{x}\right)^n q^{-nw} 
\prod_{k=0}^{n-1}\frac{q^k}{(1-q^{k+1})(\lambda-(1-\lambda)q^{k+1}
    )} \\
+ B (1-\lambda)^w &\sum_{n=0}^\infty \left(-\frac{1}{x}\right)^n
q^{-nw} \prod_{k=0}^{n-1}\frac{q^k}{(1-q^{k+1})((1-\lambda)-\lambda q^{k+1})}\;, \nonumber 
\end{align}
and so for $R_w$ via (\ref{Ransatz}).
One can then use the initial conditions (\ref{Rinitial}) to solve for
the coefficients $A$ and $B$.

Defining
\beq
\phi^{(N)}_w\left(\rho,q\right) = \sum_{n=0}^N \frac{\rho^n q^{(n-N-w+1)n}}{(q;q)_n
  (q;q)_{N-n} (\rho q;q)_n (\frac{q}{\rho};q)_{N-n}}
\eeq
where \begin{equation}
(t;q)_n=\prod_{k=0}^{n-1}(1-tq^k) 
\end{equation} is the standard
$q$-product,
after some lengthy calculations one finds
\begin{equation}
\label{Qsolnb1}
Q_w(x,y,q,a,1) = q^{w(w+1)/2}\left[(1-ax) T^{(1)}_w(x,q,\lambda) +  (1-qx
-ax(1-y^2)) T^{(2)}_w(x,q,\lambda)\right]
\end{equation}
with
\begin{multline}
\label{t1}
T^{(1)}_w (x,q,\lambda) = \frac1{1-2\lambda}\times\\ \sum_{N=0}^w
      q^{(N^2-3N)/2} (-x)^{w-N}
\left[ \frac{\lambda^w}{(1-\lambda)^{N-1}}\phi^{(N)}_w\left(\frac{1-\lambda}{\lambda},q\right)
- \frac{(1-\lambda)^w}{\lambda^{N-1}}\phi^{(N)}_w\left(\frac{\lambda}{1-\lambda},q\right)
        \right]
\end{multline}
and 
\begin{multline}
\label{t2}
T^{(2)}_w (x,q,\lambda) = -\frac1{1-2\lambda}\times\\
\sum_{N=0}^w
      q^{(N^2-N-1)/2} (-x)^{w-1-N}
\left[ \frac{\lambda^w}{(1-\lambda)^{N}}\phi^{(N)}_{w+1}\left(\frac{1-\lambda}{\lambda},q\right)
- \frac{(1-\lambda)^w}{\lambda^{N}}\phi^{(N)}_{w+1}\left(\frac{\lambda}{1-\lambda},q\right)
        \right]\;.
\end{multline}

To obtain an expression for $G_w$ one can then substitute
(\ref{Qsolnb1}) into (\ref{Qrecurrence}) to obtain an expression for
$Q_w(x,y,q,a,b)$ and then this into (\ref{Grational2}) to give
$G_w(x,y,q,a,b)$. 

\section{The infinite width limit}

For any finite length the partition function in the slit becomes equal
to the half-plane for large enough widths $w$. This means that the
generating functions $G_w$ approach $G^{hp}$, when they
converge. The half-plane solution can be found in \cite{op09}. Here we
take the limit $w\rightarrow \infty$ of the solution derived for
finite width. This gives us a more
compact expression for the denominator of the $G^{hp}$ than appears in
\cite{op09}.

After some work we find, using $\hat{\lambda}=1-\lambda$, 
\begin{equation}
Q^{hp}(x,y,q,a) = (1-ax) \left[ P^{(1)}(x,q,\lambda) +  P^{(2)}(x,q,\lambda)\right] +  (1-qx
-ax(1-y^2)) \left[ P^{(3)}(x,q,\lambda) + P^{(4)}(x,q,\lambda)\right]
\end{equation}
with
\begin{equation}
\label{p1}
P^{(1)} (x,q,\lambda) = \sum_{M=0}^\infty  \sum_{m=0}^\infty
 \frac{(-x)^M \,\lambda^{M+m} \, \hat{\lambda}^{1-m }\, q^{\frac{1}{2}M^2+Mm +m^2
   +\frac{1}{2}M -m }}{(\hat{\lambda} - \lambda) (q;q)_\infty
  (q;q)_{m} ({\lambda q}/{\hat{\lambda}};q)_m
  ({\hat{\lambda}q}/{\lambda};q)_{\infty}}\;, 
\end{equation} 
\begin{equation}
\label{p2}
P^{(2)} (x,q,\lambda)  
 = \sum_{M=0}^\infty  \sum_{m=0}^\infty
 \frac{(-x)^M \,\lambda^{1-m} \, \hat{\lambda}^{m+M}\, q^{\frac{1}{2}M^2+Mm +m^2
   +\frac{1}{2}M -m }}{(\lambda - \hat{\lambda}) (q;q)_\infty
  (q;q)_{m} ({\lambda q}/{\hat{\lambda}};q)_\infty
  ({\hat{\lambda}q}/{\lambda};q)_{m}}\;,
\end{equation}
\begin{equation}
\label{p3}
P^{(3)} (x,q,\lambda) 
 = \sum_{M=0}^\infty  \sum_{m=0}^\infty
 \frac{(-x)^M \,\lambda^{M+m+1} \, \hat{\lambda}^{-m }\, q^{\frac{1}{2}M^2+Mm +m^2
   +\frac{3}{2}M  + m }}{(\lambda - \hat{\lambda}) (q;q)_\infty
  (q;q)_{m} ({\lambda q}/{\hat{\lambda}};q)_m
  ({\hat{\lambda}q}/{\lambda};q)_{\infty}}\;,
\end{equation}
and
\begin{equation}
\label{p4}
P^{(4)} (x,q,\lambda) 
 = \sum_{M=0}^\infty  \sum_{m=0}^\infty
 \frac{(-x)^M \,\lambda^{-m} \, \hat{\lambda}^{M+m+1 }\, q^{\frac{1}{2}M^2+Mm +m^2
   +\frac{1}{2}M -m }}{(\hat{\lambda} - \lambda) (q;q)_\infty
  (q;q)_{m} ({\lambda q}/{\hat{\lambda}};q)_m
  ({\hat{\lambda}q}/{\lambda};q)_{\infty}}\;.
\end{equation}
As in \cite{op09}, we then find
\begin{equation}
G^{hp}(x,y,q,a)=\frac{Q^{hp}(x,y,q,0)}{Q^{hp}(x,y,q,a)}\;.
\end{equation}
While this expression is equivalent to that found in \cite{op09}, we note that 
its structure is fundamentally different.

\section{Mapping to Motzkin paths polymer model}

\begin{figure}[ht!]
\begin{center}
\includegraphics[width=0.3\textwidth]{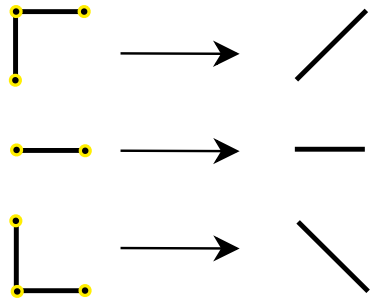}
\caption{The mapping from RSOS paths to Motzkin paths.}
\label{figure3}
\end{center}
\end{figure}

\begin{figure}[ht!]
\begin{center}
\includegraphics[width=0.7\textwidth]{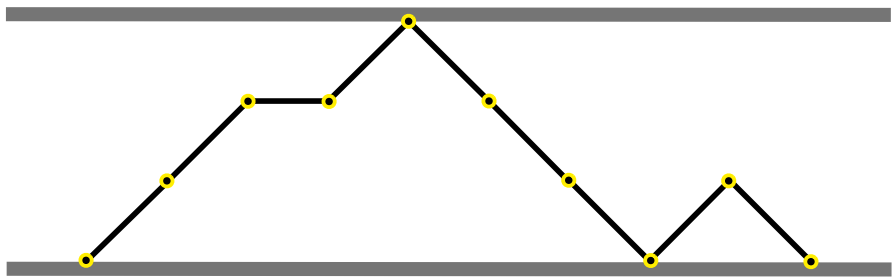}
\caption{The mapping of RSOS configuration in Figure 1 to a Motzkin path.}
\label{figure4}
\end{center}
\end{figure}

As foreshadowed in the Introduction there is a mapping between RSOS
configurations and Motzkin paths. Starting at the leftmost site one
uses the mapping in Figure~\ref{figure3} to construct a Motzkin path
from an RSOS configuration.
For example, the mapping of configuration in Figure 1 can be seen in Figure~\ref{figure4}.
This implies that the generating function we have found is also that
of of Motzkin paths in a slit weighted by area (which can be used to
model an osmotic pressure \cite{bop09}) an interaction with both
walls. In this way the current work is a generalisation of that found
in \cite{bop09}.

\section{Conclusion}

In this paper we have presented a solution to the linear RSOS model in
a slit with field and wall interaction terms in the energy. In
particular we have evaluated the generating function and demonstarted
that its limit is the half-plane solution found earlier. The numerator
and denominator polynomials of the slit generating function are novel
$q$-orthogonal polynomials associated with the continued fraction
expansion of the half-plane solution.

\section*{Acknowledgements}

Financial support from the Australian Research Council via its support
for the Centre of Excellence for Mathematics and Statistics of Complex
Systems is gratefully acknowledged by the authors. A L Owczarek thanks the
School of Mathematical Sciences, Queen Mary, University of London for
hospitality.


\begin{thebibliography}{10}

\bibitem{te52}
H. N. V. Temperley,
Proc. Camb. Phil. Soc. {\bf 48}, 638 (1952).

\bibitem{di88} 
S. Dietrich, in {\em Phase Transitions and Critical Phenomena},
Vol. 12, ed.\ by C. Domb and J. L. Lebowitz, (Academic Press, London, 1988).

\bibitem{prsv89} 
V. Privman  and N. M. \v Svraki\'c,
{\em Lecture Notes in Physics} {\bf 338} (Springer--Verlag, Berlin,
1989).

\bibitem{owczarek93a}
{A. L. Owczarek and T. Prellberg, 
J. Stat. Phys. {\bf 70} 1175 (1993)}

\bibitem{borw05}
R. Brak, A.L. Owczarek, A. Rechnitzer and S.G. Whittington
J. Phys. A: Math. Gen., {\bf 38}: 4309, (2005).


\bibitem{bop09}
A. L. Owczarek and T. Prellberg,
{A simple model of a vesicle drop in a confined geometry}, 
submitted to JSTAT (2010).

\bibitem{op09}
A. L. Owczarek and T. Prellberg,
J. Phys. A: Math. Gen., {\bf 42}: 495003, (2009).

\bibitem{birw07}
R. Brak, G.K. Iliev, A. Rechnitzer and S.G. Whittington
J. Phys. A: Math. Theor., {\bf 40}, 4415, (2007).

\bibitem{andrews1999a-a}
G.~E. Andrews, R.~Askey, and R.~Roy,
\newblock volume~71 of {\em Encyclopedia of Mathematics and its Applications},
\newblock Cambridge University Press, Cambridge, 1999.


\end{thebibliography}
\end{document}